\documentclass[ras]{agutex}
\usepackage{graphicx, epstopdf, color, bm, amsmath}

\authorrunninghead{LOI ET AL.}

\titlerunninghead{NEW ANGLE FOR PROBING FAIS WITH THE MWA}

\begin{document}

%% ------------------------------------------------------------------------ %%
%
%  TITLE
%
%% ------------------------------------------------------------------------ %%

\title{A new angle for probing field-aligned irregularities with the Murchison Widefield Array}
%
% e.g., \title{Terrestrial ring current:
% Origin, formation, and decay $\alpha\beta\Gamma\Delta$}
%

%% ------------------------------------------------------------------------ %%
%
%  AUTHORS AND AFFILIATIONS
%
%% ------------------------------------------------------------------------ %%

%Use \author{\altaffilmark{}} and \altaffiltext{}

\authors{Shyeh Tjing Loi,\altaffilmark{1,2,3}
Tara Murphy,\altaffilmark{1,2}
Iver H.~Cairns,\altaffilmark{4}
Cathryn M.~Trott,\altaffilmark{2,5}
Natasha Hurley-Walker,\altaffilmark{5}
Lu Feng,\altaffilmark{6}
Paul J.~Hancock,\altaffilmark{2,5}
David L.~Kaplan\altaffilmark{7}
}

\altaffiltext{1}{Sydney Institute for Astronomy, School of Physics, University of Sydney, New South Wales, Australia}

\altaffiltext{2}{ARC Centre of Excellence for All-sky Astrophysics, Sydney, New South Wales, Australia}

\altaffiltext{3}{Department of Applied Mathematics and Theoretical Physics, Centre for Mathematical Sciences, University of Cambridge, Cambridge, United Kingdom}

\altaffiltext{4}{School of Physics, University of Sydney, New South Wales, Australia}

\altaffiltext{5}{International Centre for Radio Astronomy Research, Curtin University, Bentley, Western Australia, Australia}

\altaffiltext{6}{Kavli Institute for Astrophysics and Space Research, Massachusetts Institute of Technology, Cambridge, Massachusetts, USA}

\altaffiltext{7}{Department of Physics, University of Wisconsin-Milwaukee, Milwaukee, USA}

% \altaffilmark will produce footnote;
% matching \altaffiltext will appear at bottom of page.

% \authors{A. B. Smith,\altaffilmark{1}
% Eric Brown,\altaffilmark{1,2} Rick Williams,\altaffilmark{3}
% John B. McDougall\altaffilmark{4}, and S. Visconti\altaffilmark{5}}

%\altaffiltext{1}{Department of Hydrology and Water Resources,
%University of Arizona, Tucson, Arizona, USA.}

%\altaffiltext{2}{Department of Geography, Ohio State University,
%Columbus, Ohio, USA.}

%\altaffiltext{3}{Department of Space Sciences, University of
%Michigan, Ann Arbor, Michigan, USA.}

%\altaffiltext{4}{Division of Hydrologic Sciences, Desert Research
%Institute, Reno, Nevada, USA.}

%\altaffiltext{5}{Dipartimento di Idraulica, Trasporti ed
%Infrastrutture Civili, Politecnico di Torino, Turin, Italy.}

%% ------------------------------------------------------------------------ %%
%
%  ABSTRACT
%
%% ------------------------------------------------------------------------ %%

% >> Do NOT include any \begin...\end commands within
% >> the body of the abstract.

\begin{abstract}
Electron density irregularities in the ionosphere are known to be magnetically anisotropic, preferentially elongated along the lines of force. While many studies of their morphology have been undertaken by topside sounding and whistler measurements, it is only recently that detailed regional-scale reconstructions have become possible, enabled by the advent of widefield radio telescopes. Here we present a new approach for visualising and studying field-aligned irregularities (FAIs), which involves transforming interferometric measurements of TEC gradients onto a magnetic shell tangent plane. This removes the perspective distortion associated with the oblique viewing angle of the irregularities from the ground, facilitating the decomposition of dynamics along and across magnetic field lines. We apply this transformation to the dataset of \citet{Loi2015_mn2e}, obtained on 15 October 2013 by the Murchison Widefield Array (MWA) radio telescope and displaying prominent FAIs. We study these FAIs in the new reference frame, quantifying field-aligned and field-transverse behaviour, examining time and altitude dependencies, and extending the analysis to FAIs on sub-array scales. We show that the inclination of the plane can be derived solely from the data, and verify that the best-fit value is consistent with the known magnetic inclination. The ability of the model to concentrate the fluctuations along a single spatial direction may find practical application to future calibration strategies for widefield interferometry, by providing a compact representation of FAI-induced distortions.
\end{abstract}

%% ------------------------------------------------------------------------ %%
%
%  BEGIN ARTICLE
%
%% ------------------------------------------------------------------------ %%

% The body of the article must start with a \begin{article} command
%
% \end{article} must follow the references section, before the figures
%  and tables.

\begin{article}

%% ------------------------------------------------------------------------ %%
%
%  TEXT
%
%% ------------------------------------------------------------------------ %%

\section{Introduction}
\subsection{Plasma Density Irregularities}
The Earth's plasma environment abounds with density irregularities, on scales ranging from those of neutral atmosphere waves (100--1000\,km) down to ion and electron gyroradii (0.1--1\,m) \citep{Yeh1974, Booker1979, Fung2000, Akmaev2011, Wang2011, Nicolls2014}. They tend to be elongated along the geomagnetic field lines owing to large electron mobilities in this direction, and are broadly referred to as field-aligned irregularities (FAIs) \citep[e.g.][]{Sonwalkar2006, Makela2012}. Acoustic-gravity waves (AGWs), which couple to charged species via collisions, are believed to be the source of energy driving the formation of smaller-scale density structures, for example via the spatial resonance mechanism \citep{Whitehead1971, Klostermeyer1978} or upon passage through a critical layer \citep{Booker1967, Staquet2002}, with further cascades possible through various plasma instabilities and nonlinear processes \citep{Perkins1973, Fejer1980, Maruyama1990, Nicolls2005}. Observational evidence for the association between AGWs and FAIs was recently presented by \citet{Sun2015} and \citet{Loi2016}.

FAIs of suitable dimensions and density contrasts are capable of ducting and guiding VLF to HF radio waves, a property that enabled their initial discovery \citep{Storey1953} and subsequent investigations \citep{Calvert1969, Angerami1970, Singh1998, Darrouzet2009}. Studies of their morphology have mostly relied on theoretical knowledge about electromagnetic (EM) wave trapping and guidance by plasma density structures \citep{Smith1960, Calvert1995}, aided by ray tracing calculations to interpret received signal patterns \citep{Haselgrove1954, Smith1961, Dyson1967, Hayakawa1978, Fung2005, Kulkarni2008}. Sheet-like structures extended in the east-west direction have been posited as one interpretation of topside sounder data \citep{Muldrew1963, Dyson1967}, possibly occurring in uniformly spaced onion-like layers \citep{Gross1984}. 

Numerical studies suggest that tubular structures can also be effective waveguides \citep{Smith1961, Platt1989}. The existence of tubular structures with inferred widths of 10--100\,km has been established through satellite \textit{in situ} measurements \citep{Angerami1970, Sonwalkar1994, Decreau2005}, satellite remote sensing \citep{Fung2003, Darrouzet2009, Woodroffe2013}, ground-based whistler spectrograms \citep{Hayakawa1978, Singh1998, Altaf2013}, and ground-based radio interferometric arrays \citep{Jacobson1993, Hoogeveen1997, Helmboldt2012b}. Recent interferometric observations support the idea that tubular FAIs may in some instances be confined to a narrow sheet-like layer on a single magnetic shell \citep{Loi2015_mn2e, Loi2015a_mn2e}. Subtler details such as field-aligned density profiles have been measured through VLF whistler observations \citep{Carpenter1964, Lichtenberger2009, Lichtenberger2013}, ULF field-line resonances \citep{Menk1999}, and topside sounding \citep{Dyson1978, Tu2006}. 

The growth and decay rates of FAIs depend on many factors, including the electrical conductivity, electric and magnetic fields, ambient densities and density gradients, collisional frequencies and drift velocities \citep{FarleyJr1963, Fejer1980, Ossakow1981, Sojka1998, Fejer1999}. Numerical and experimental results suggest that growth due to various instabilities may occur on timescales of minutes to tens of minutes \citep{Park1971a, Sojka1998, Carter2014}, and decay by cross-field diffusion on timescales of hours to days \citep{Thomson1978, Singh1999, Singh2013}.

\begin{figure*}
  \centering
  \includegraphics[width=0.65\textwidth]{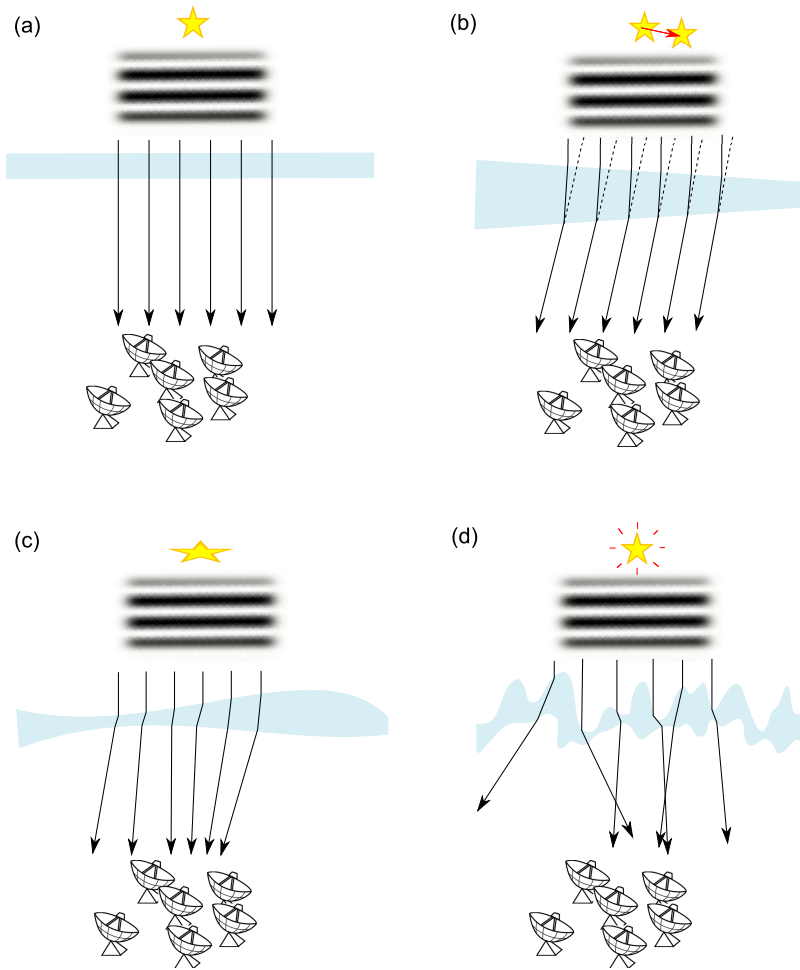}
  \caption{A cartoon illustration of the effects of phase distortions induced by the ionosphere. In the case of a smooth ionosphere (a), there is no change in the differential phase. In the linearly varying case (b), a simple angular shift results. Higher order spatial variations (c) cause shape distortion. For sufficiently small irregularities (d) a non-linear diffractive regime is reached, resulting in scintillation.}
  \label{fig:propagation_effects}
\end{figure*}

\subsection{Radio Interferometry}\label{sec:interferometry}
Many radio telescopes are interferometric arrays that measure the phase difference of incoming EM waves between pairs of receivers \citep{Thompson2001}. Since the refractive index of a plasma is electron density-dependent, interferometers can detect density gradients through phase fluctuations. In an interferometric image, these manifest as shifts in the angular position of a radio source and/or broadening of the angular extent of a source, and may also be accompanied by intensity fluctuations \citep{Hamaker1978, Bougeret1981, Spoelstra1984, Jacobson1992a, Cotton2004, Cohen2009}. It is in the interest of those using radio telescopes for astronomical observations to model and subtract these distortions from (i.e.~calibrate) the data. The development of ionospheric calibration strategies for newly emergent widefield interferometers operating in the VHF band is an ongoing challenge, with a variety of approaches being trialled \citep{Cotton2004, Intema2009, Arora2015_mn2e, vanWeeren2016}.

Bulk angular shifts are caused by density gradients whose scale lengths greatly exceed the physical size of the interferometric array, forming a wedge that collectively tilts the wavefront normals arriving at all baselines \citep{Thompson2001}. The associated displacement $\Delta\theta$ (in radians) is directly proportional to the transverse TEC gradient $\nabla_\perp$TEC (in el\,m$^{-3}$) and inversely proportional to the observing frequency $\nu$ (in Hz) according to
%\begin{linenomath*}
\begin{equation}
  \Delta\theta \approx -\frac{40.3}{\nu^2} \nabla_\perp \text{TEC} \:. \label{eq:displacement}
\end{equation}
%\end{linenomath*}
Angular broadening can be conceived as the result of superposing multiple wavefront normals, which might occur if irregularity scale sizes fall below the size of the array (different baselines experience different phase deviations), or if integration times exceed the timescale of variation in the plasma along the line of sight \citep{Spoelstra1984, Lonsdale2005, Kassim2007}. Intensity variations can arise from focusing and defocusing of the incoming rays, becoming extreme (modulation indices of order unity) in the diffraction-dominated limit \citep{Meyer-Vernet1980, Booker1981, Spoelstra1985}. These effects are illustrated in the cartoon in Figure \ref{fig:propagation_effects}. Note that a perfectly uniform TEC screen cannot be detected by an interferometer, since it is only the phase difference and not the absolute phase that is measured on a baseline (it is in principle possible to measure the absolute TEC through Faraday rotation, but the precision of this is lower).

To perform the phase measurements the telescope must have a source of back-illuminating radio waves. These can be naturally occurring, e.g.~solar active regions \citep{Bougeret1981, Mercier1986} or cosmic radio sources \citep{Hamaker1978, Jacobson1993, Spoelstra1997, Helmboldt2014a, Loi2015a_mn2e}, or of man-made origin such as satellite beacons \citep{VanVelthoven1990_phd, Jacobson1996, Hoogeveen1997a}. Celestial sources of radio emission, though faint in comparison to solar or man-made sources, are stable in output and detectable at a rate of about one per square degree on the sky, for a confusion-limited interferometer with 2\,arcmin angular resolution operating in the VHF band. The exploitation of their abundance to construct spatially detailed, regional-scale TEC gradient maps has been made possible by the recent development of sensitive, widefield radio telescopes \citep{Loi2015a_mn2e, Loi2015b_mn2e}.

Detections of tubular FAIs by interferometric arrays were previously made using the Westerbork Radio Synthesis Telescope \citep{VanVelthoven1990_phd}, the Very Large Array (VLA) \citep{Jacobson1992, Jacobson1993, Hoogeveen1997, Helmboldt2012c}, and the Los Alamos interferometer \citep{Jacobson1996, Hoogeveen1997a}. Increasingly sophisticated analysis methods for the VLA, designed to cope with its sparse spatial sampling, have been under recent development \citep{Cohen2009, Coker2009, Helmboldt2012, Helmboldt2012b, Helmboldt2014a}. The sparseness with which traditional interferometers sample the TEC distribution is contrasted by the breadth and detail achieved by new-generation instruments such as the Murchison Widefield Array (MWA) \citep{Lonsdale2009_mn2e, Bowman2013_mn2e, Tingay2013_mn2e}, which permit rich visualisations of FAIs and travelling disturbances over wide fields of view \citep{Loi2015_mn2e, Loi2016}. Tubular FAIs have been observed in half or more of nighttime MWA data, to varying degrees of prominence \citep{Loi2015a_mn2e}. Their high occurrence rate over the site of the MWA suggests that they must be accounted for if a thorough calibration of ionospheric effects is desired. However, their steep inclination with respect to the ground implies that constant-altitude screen models might be unsuitable and that a more flexible model is needed to capture their properties.

\subsection{This Work}
We present a new method for visualising and studying FAIs, designed for widefield interferometers, that involves mapping TEC gradients measured from radio synthesis images onto a plane tangent to a magnetic shell. This transformation removes the perspective distortion associated with inclined magnetic field lines, providing a physically meaningful reference frame in which to inspect their properties. We apply this technique to the dataset previously analysed by \citet{Loi2015_mn2e}, known to exhibit prominent FAIs. We also extend the analysis to sub-array scales, something which has not been previously done with MWA data, by examining angular broadening effects.

The paper is structured as follows. In Section \ref{sec:model} we introduce the model. In Section \ref{sec:methods} we introduce the MWA observations, explain the analysis approaches for large- and small-scale irregularities, and demonstrate how the best-fit inclination may be obtained. Our results are presented in Sections \ref{sec:superMWA} and \ref{sec:subMWA} for large- and small-scale irregularities, respectively. We conclude in Section \ref{sec:conclusion}.

\section{The Inclined Plane Model}\label{sec:model}
The quantity measured by an interferometer through the angular displacement of an unresolved (point-like) radio source is $\nabla_\perp$TEC, the gradient of the TEC transverse to the line of sight to the source. Because it involves an absolute reference length scale (the spacing between receivers), this gradient is independent of the distance to the irregularities causing the phase perturbation. It is a two-dimensional vector, lying in the plane perpendicular to the line of sight. In addition, the angular position of the origin of this vector is known, given by the angular position of the source, also a two-component vector. For a celestial source, this is usually specified in terms of the right ascension $\alpha$ and declination $\delta$ (not to be confused with magnetic declination), which are coordinates fixed with respect to the celestial sphere. There are in total thus four independent input quantities: $\alpha$, $\delta$, $\partial_\alpha$TEC and $\partial_\delta$TEC, where the partial derivative shorthand $\partial_X$ denotes the gradient along direction $X$.

\begin{figure}
  \centering
  \includegraphics[width=0.5\textwidth]{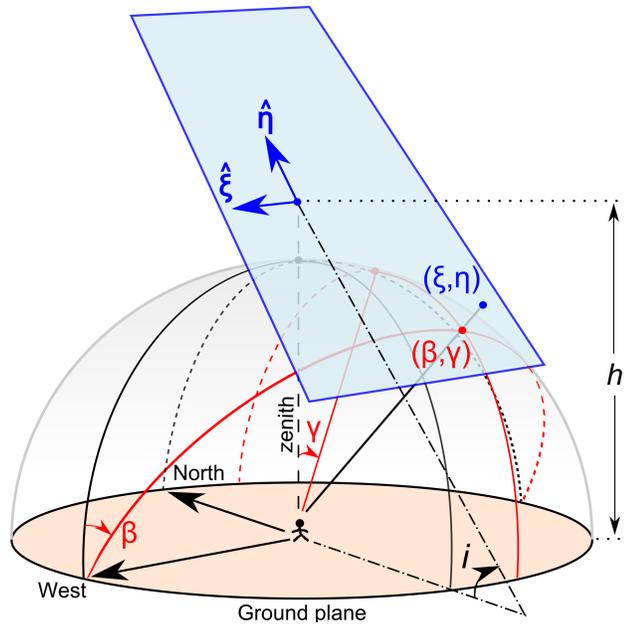}
  \caption{Diagram showing the geometry of the model and how the $(\beta,\gamma)$ and $(\xi,\eta)$ coordinate systems are defined with respect to the ground observer. Cardinal directions are as marked. The $(\beta,\gamma)$ coordinates, shown in red, are angular coordinates for an imaginary sphere centred on the observer (note that the finite radius of the sphere as drawn in this diagram is solely for visualisation purposes, and otherwise has no meaning). These are the same $(\beta,\gamma)$ coordinates as described in \citet{Loi2015a_mn2e}. The $(\xi,\eta)$ coordinates, shown in blue, are Cartesian coordinates describing position on the inclined plane. The origin of this system is defined to be the point on the plane directly overhead of the observer. The two parameters describing the plane, which are the inclination angle $i$ and zenith altitude $h$, are defined as shown. In the current model the normal vector to the plane is assumed to have zero (cardinal) east-west component.}
  \label{fig:geometry}
\end{figure}

Geometric arguments presented by \citet{Loi2015_mn2e} and \citet{Loi2015a_mn2e} indicate that FAIs observed over the MWA site are often confined to a narrow layer, i.e.~small $L$-value \citep{McIlwain1961} range, though this value of $L$ may differ between observations. This motivates the introduction of a model that maps the four observables onto such a surface. In this work, we consider the lowest-order approximation to achieve this, namely an inclined 2D plane (i.e.~a flat screen) tangential to the $L$-shell on which the irregularities reside. We assume that all phase perturbations are taken up by this screen, whose thickness is negligible. Clearly this is only valid locally since the curvature of field lines is neglected. Given that the scale of curvature is of order the radius of the Earth (several thousand kilometres) and that the measurements probe a region several hundred kilometres across, we expect this approximation to be reasonable.

Although a general 2D plane in 3D space requires three parameters for a full description, one of these can be eliminated by enforcing the east-west component of the normal vector to be zero (N.B.~at the site of the MWA, geomagnetic and geographic east-west coincide to within a degree and so their difference can be conveniently neglected; see Appendix \ref{sec:trafo} for how to handle the more general case). This is equivalent to assuming axisymmetry of the Earth's magnetic field, a good approximation for the inner magnetosphere where the field is close to being dipolar. This leaves two independent parameters, which we define to be $h$, the altitude of the plane at the point directly above the MWA, and $i$, the inclination angle (measured positive for a plane sloping upwards towards the north). These two quantities are illustrated in Figure \ref{fig:geometry}. Suitable values for $h$ and $i$ can in fact be derived solely from the data without any knowledge of the geomagnetic field: $h$ can be estimated via a parallax technique \citep{Loi2015_mn2e} and $i$ can be optimised by demanding that it maximises the ``parallelness'' of resultant features (quantitative details in Section \ref{sec:inclination}). 

\begin{figure*}
  \centering
  \includegraphics[width=0.9\textwidth]{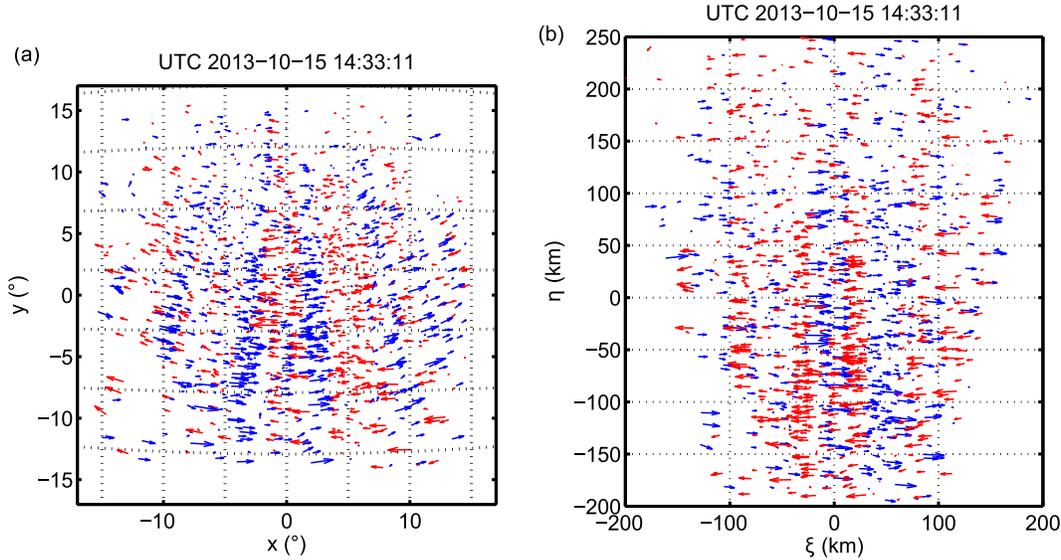}
  \caption{A comparison of the electron density gradient vector field from a particular snapshot computed with respect to (a) a Sanson-Flamsteed projection of the $(\beta,\gamma)$ coordinates, and (b) the new $(\xi,\eta)$ coordinates. The parameters used are $h = 570$\,km and $i = 59^\circ$. Note that in both cases the sky is into the page, so that the vertical axis points north and the horizontal axis points west. Horizontal and vertical dotted lines in (a) correspond to lines of constant $\beta$ and $\gamma$, while those in (b) correspond to lines of constant $\eta$ and $\xi$, respectively. In (a), the relationship between $(x,y)$ and $(\beta,\gamma)$ is given by equation (3) of \citet{Loi2015a_mn2e}. It is to be noted that the $(x,y)$ and $(\beta,\gamma)$ systems nearly coincide (the dotted lines are very close to being $x$ and $y$ gridlines). Arrows lengths are directly proportional to the magnitude of the gradient vector and scaled for clarity, such that a $1^\circ$-long arrow in (a) represents a gradient of $2.9 \times 10^{14}$\,el\,m$^{-2}$\,km$^{-1}$ while a 10-km long arrow in (b) represents a gradient of $1.4 \times 10^{14}$\,el\,m$^{-2}$\,km$^{-1}$. Blue and red colours are an aid to visualisation and denote arrows with positive and negative horizontal components, respectively.}
  \label{fig:arrow}
\end{figure*}

To describe position on the plane itself, we define an orthogonal basis $(\hat{\boldsymbol{\xi}}, \hat{\boldsymbol{\eta}})$ where the $\hat{\boldsymbol{\xi}}$-axis points west and the $\hat{\boldsymbol{\eta}}$-axis points north (see Figure \ref{fig:geometry}). The equations for how to map celestial coordinates $(\alpha, \delta)$ to screen coordinates $(\xi,\eta)$, which make use of the intermediary $(\beta,\gamma)$ system introduced by \citet{Loi2015a_mn2e}, are given in Appendix \ref{sec:trafo}. Note that the $(\xi,\eta)$ coordinates are not geographically absolute but are defined with respect to the observing location, with $\xi = \eta = 0$ corresponding to the observer's zenith.

As explained earlier in Section \ref{sec:interferometry}, an interferometer is insensitive to the zeroth-order (i.e.~constant offset) component of the TEC distribution. This implies that if density irregularities are confined to a thin layer bounded above and below by smooth plasma, then this is observationally equivalent to a thin plasma screen bounded above and below by vacuum. In the context of our model, we thus find it natural to associate a surface density rather than a column density to points on the screen (although given that the two have equivalent units, the difference is mainly conceptual). We denote this surface density by $\Sigma$, a scalar function of $\xi$ and $\eta$. The gradient of $\Sigma$ with respect to the screen can be determined by a suitable transformation of the $\nabla_\perp$TEC vector. For our simple model, it is possible to obtain explicit expressions for $\partial_\xi\Sigma$ and $\partial_\eta\Sigma$ in terms of the four input quantities. These and further details are contained in Appendix \ref{sec:trafo}.

The effect of the transformation from angular to screen coordinates for an arbitrary snapshot from the dataset of \citet{Loi2015_mn2e} is demonstrated in Figure \ref{fig:arrow}. The coordinate system in Figure \ref{fig:arrow}a is the one which has been used for visualisation in previous MWA work, and corresponds to the view one might have looking up at the structures from the ground (see \citet{Loi2015a_mn2e} for a qualitative overview and Appendix \ref{sec:trafo} of the current work for quantitative details). On the other hand, Figure \ref{fig:arrow} is the view one might have from a vantage point looking ``face-on'' at the surface on which the irregularities reside. We can visually identify two consequences of the transformation: (i) there is an ``unskewing'' of apparently convergent features into a much more parallel configuration, and (ii) the gradient vectors turn out to be nearly parallel to the $\xi$ axis. While the first property is to be expected from the way in which $i$ was chosen (to maximise the ``parallelness'' of features; details in Section \ref{sec:inclination}), the second property is an independent verification that the model is physically reasonable (larger density gradients can be sustained across than along field lines).

\begin{figure*}
  \centering
  \includegraphics[width=0.9\textwidth]{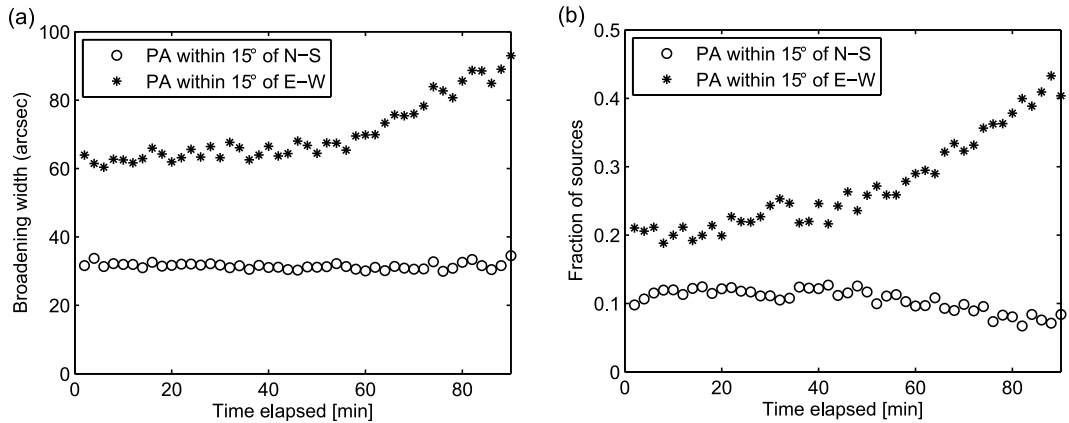}
  \caption{(a) The broadening width (quadrature difference of major and minor axes) as a function of time, for sources whose position angles are aligned with the $\xi$ and $\eta$ axes. The alignment criterion has been taken to be a position angle within $15^\circ$ of the respective axis. (b) The fractions of sources selected by these restrictions on orientation, as a function of time. This analysis makes use of the 4-s cadence data to minimise the spatial extent of the region sampled by the drifting sources. Only sources of signal-to-noise ratio greater than 10 have been included.}
  \label{fig:broad_vs_time}
\end{figure*}

\section{Methods}\label{sec:methods}
\subsection{Instrument, Observations and Data}
The MWA telescope is an interferometric array sited at a mid-latitude location in Western Australia, at a geographic latitude of $26.7^\circ$S and a geomagnetic latitude of $38.6^\circ$S ($L = 1.6$). It operates at frequencies between 80 and 300\,MHz (in the VHF band). Its 128 receiving elements (``tiles'') are spread over a total region about 3\,km wide on the ground but are centrally condensed, with most (112 out of 128) tiles lying within 0.75\,km of the core. The MWA therefore behaves as a point detector for structures greatly exceeding 1\,km, and a distributed array for structures on much smaller scales. The instantaneous angular field of view (FOV) of the MWA is $30^\circ$, subtending a physical distance of about 300\,km at 600\,km altitude.

The dataset considered in this paper is the same as the one discussed by \citet{Loi2015_mn2e}. It was obtained during mildy disturbed geomagnetic conditions ($K_p = 2$) within the recovery phase of a moderate (minimum $Dst = -45$\,nT, 24-hr maximum $K_p = 4$) geomagnetic storm on 15 October 2013. The data were recorded at 183\,MHz (30.72\,MHz instantaneous bandwidth) over a 1.5-hr long interval between 1346 and 1517 UTC (2146--2317 Australian Western Standard Time). Prominent FAIs are present over the whole FOV during this period. Their characteristic altitude was established by \citet{Loi2015_mn2e} to be $570 \pm 40$\,km.

The interval comprises 46 blocks of data spaced by 2\,min and recorded over 112\,s each. However, the interferometric visibilities (i.e.~complex voltages) were in fact captured at 0.5\,s time resolution. While most of the analysis here uses images formed by integrating over the 112\,s of data in each 2-min block, a portion of this work (Section \ref{sec:subMWA}) makes use of images formed at a much higher cadence of 4\,s. This allows short-term dynamics to be elucidated at the expense of diminished image sensitivity.

\subsection{Probing Large-scale Structure}
As discussed earlier, the size of the array determines how phase fluctuations of various scales manifest in the data. Hereafter we define ``large-scale'' to mean structures larger than the diameter of the MWA (``super-MWA'' scales). Practically speaking, these are the structures that can be probed through the displacements they induce in the positions of unresolved radio sources. The method for extracting these angular displacements in the current work is identical to that described in previous MWA work \citep{Loi2015_mn2e, Loi2015a_mn2e, Loi2015b_mn2e, Loi2016}, to which we refer the reader for further details. 

Briefly, this involved first identifying candidate radio sources by searching for intensity maxima in the images above a given noise threshold, cross-matching them with a published astronomical database (here the National Radio Astronomy Observatory VLA Sky Survey, NVSS; \citet{Condon1998}) and retaining only those with counterparts. The displacement vector $\Delta\theta$ associated with an NVSS source appearing in a certain snapshot was taken to be the difference between its position measured in that snapshot and its time-averaged position. The $\Delta\theta(\alpha,\delta)$ vectors were then mapped to $\nabla\Sigma(\xi,\eta)$ vectors by the transformation steps described in Section \ref{sec:model} and Appendix \ref{sec:trafo}. Results pertaining to large-scale structure are discussed in Section \ref{sec:superMWA}.

\subsection{Probing Small-scale Structure}
We define ``small-scale'' here to refer to structures on sub-MWA scales. These fluctuations are responsible for the apparent broadening of otherwise unresolved radio sources; one can envisage this as arising from the simultaneous arrival of wavefronts with different orientations over different parts of the array. The spread of angular offset vectors causes the source to appear blurred. The amount and orientation of the broadening relate to the spread in wavefront orientations, which in turn arises from the spread in $\nabla_\perp$TEC on sub-array scales. The ``broadening vector'' (defined in more detail below) associated with this process therefore obeys the same transformation rules as the $\nabla_\perp$TEC vectors measured through bulk angular displacements. 

To probe structures in the sub-MWA regime we made use of images formed at 4-s rather than 2-min cadence, to minimise the size of the region sampled towards each source. The drift of celestial sight lines through the screen occurs at a speed of $\sim$40\,m\,s$^{-1}$ at 600\,km altitude, meaning that each patch of sight lines sweeps out an area roughly three times larger in a 2-min compared to a 4-s image, assuming an effective MWA diameter of 1.5\,km ($4.8 \times 1.5$\,km$^2$, versus $1.7 \times 1.5$\,km$^2$).

The software package used for automated source extraction (\textsc{Aegean}; \citet{Hancock2012}) operates by fitting 2D elliptical Gaussians to intensity maxima. Besides the best-fit coordinates of the centre of the Gaussian, \textsc{Aegean} also reports the major axis $b_\text{maj}$, minor axis $b_\text{min}$, and position angle $p$ (measured east of north). In the absence of any scatter broadening, the width of a source in an image is given by the width of the synthesised beam (response function), which for the MWA at 183\,MHz is about 2\,arcmin. The MWA synthesised beam is approximately a circular Gaussian, but the exact shape and size can vary over time if certain tiles/baselines are temporarily excised due to interference or hardware malfunctions (largely an automated process). The broadening due to finite angular resolution is distinct from scatter broadening.

Our approach to isolating the scatter broadening contribution was to take the quadrature difference between the major and minor axes fitted for the Gaussian, which we call the \textit{broadening width}. This leads to our definition of the \textit{broadening vector} $\mathbf{b}$ as the vector oriented parallel to the major axis whose associated magnitude is the broadening width:
%\begin{linenomath*}
\begin{equation}
  \mathbf{b} = \pm \sqrt{b_\text{maj}^2 - b_\text{min}^2} \left(
    \frac{\sin p}{\cos \delta} \:, \cos p \right) \:,
  \label{eq:broadening}
\end{equation}
%\end{linenomath*}
given with respect to the $(\alpha,\delta)$ basis. The $\pm$ sign reflects the fact that $\mathbf{b}$ is a spin-2 vector (since the orientation of the major axis is only defined modulo 180$^\circ$). The broadening vectors take the role of the angular displacement vectors $\Delta\theta$ described in Section \ref{sec:model} and Appendix \ref{sec:trafo} as the measure of TEC fluctuation. We subjected the $\mathbf{b}$ vectors to an identical sequence of transformations to obtain the sub-MWA version of ``$\nabla\Sigma$'', which is more closely related to the second spatial derivative of $\Sigma$.

\begin{figure}
  \centering
  \includegraphics[width=0.5\textwidth]{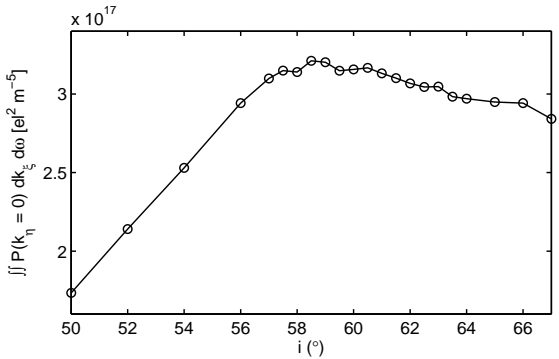}
  \caption{The degree of ``$\xi$-parallelness'' of features, measured through a spatial Fourier transform (details in text), as a function of the plane inclination $i$. This maximises at around $i = 59^\circ$, which is consistent with the known inclination of geomagnetic field lines within the MWA FOV (see text for discussion). A value of $i = 59^\circ$ has been adopted for all analyses in this work.}
  \label{fig:optimal_inc}
\end{figure}

Note that the manner in which we have quantified the scatter broadening contribution assumes that the broadening occurs primarily along a selected direction, and is therefore a conservative lower bound on the actual amount of broadening. Physically we expect density irregularities below $\sim$1\,km scales to be highly anisotropic, because these scales are less than the mean free path of neutrals in the thermosphere (several kilometres) \citep{Jacchia1977} and shaped by magnetic forces rather than collisions. We thus expect small-scale irregularities to be strongly field-aligned, producing broadening in the direction perpendicular to their elongation (cf.~a diffraction grating). Indeed, the data exhibit both a larger fraction of sources broadened along $\xi$ and a larger average broadening width in this direction, compared to $\eta$ (Figure \ref{fig:broad_vs_time}).

Referring to the average broadening width for sources whose $\mathbf{b}$ vectors lie within $15^\circ$ of the $\xi$ and $\eta$ axes as $w_\xi$ and $w_\eta$ respectively, we also see from Figure \ref{fig:broad_vs_time} that $w_\xi$ grows with time. This is closely reminiscent of the growth of the larger-scale structures apparent in figure 3c of \citet{Loi2015_mn2e}. In contrast, $w_\eta$ exhibits no such growth. Noting that $w_\eta$ turns out to be comparable to the error on $b_\text{maj}$ and $b_\text{min}$ quoted by \textsc{Aegean}, we can regard this as a measure of the noise floor. Subtracting this in quadrature from $w_\xi$, we thus arrive at the following conservative estimate of the spread in $\nabla\Sigma$ due to small-scale FAIs:
%\begin{linenomath*}
\begin{equation}
  \sigma(\nabla\Sigma) = \frac{1}{2} \sqrt{w_\xi^2 - w_\eta^2} \:. \label{eq:broadwid}
\end{equation}
%\end{linenomath*}
The factor of 1/2 reflects the fact that $w_\xi$ and $w_\eta$ are major axis-related quantities while $\sigma$ is a root-mean-square (RMS) quantity, which is related to the semimajor axis of a Gaussian distribution. Unit conversion factors have been omitted for clarity. The results for small-scale structure are presented in Section \ref{sec:subMWA}.

\begin{figure*}
  \centering
  \includegraphics[width=0.9\textwidth]{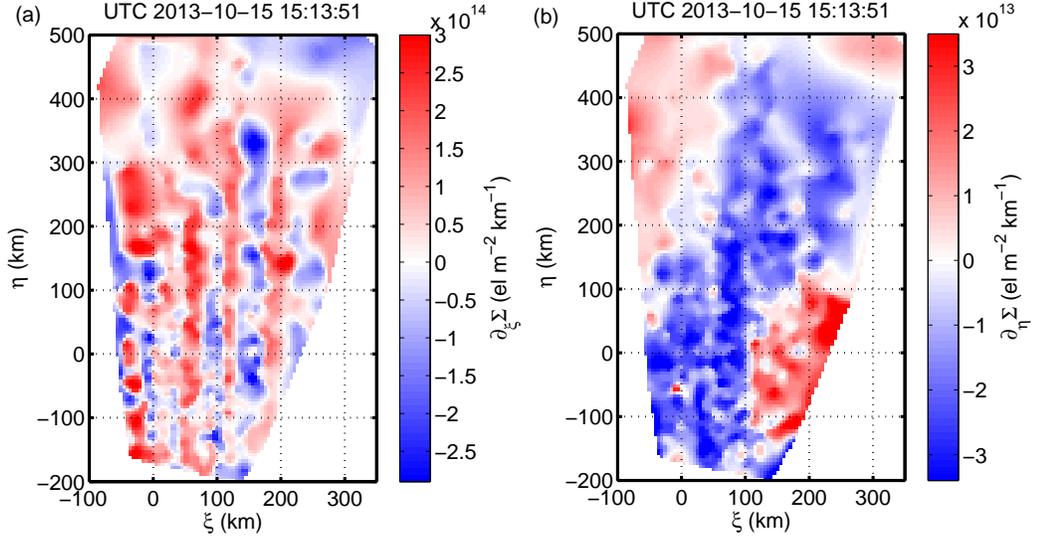}
  \caption{The (a) $\xi$ and (b) $\eta$ components of the $\nabla\Sigma$ vector field, for a particular snapshot. Red and blue correspond to positive and negative gradients, and white corresponds either to near-zero values or regions outside the FOV (i.e.~unmeasured values). These have been generated by first forming a natural-neighbour interpolant over the data, resampling over a 5-km uniform grid and then smoothing the result with a 3$\times$3-box median filter. The shorthand $\partial_X \equiv \partial/\partial X$ has been used to denote a partial derivative. See Movie S1 for an animation of the full dataset.}
  \label{fig:shading_xieta}
\end{figure*}

\subsection{Choosing the Height and Inclination}\label{sec:inclination}
In Section \ref{sec:model} we introduced the two free parameters $h$ and $i$, which are inputs to the model and illustrated in Figure \ref{fig:geometry}. The best-fit average height of the irregularities in the 15 October 2013 dataset was determined previously by \citet{Loi2015_mn2e} to be $570 \pm 40$\,km, and so the value of $h$ we adopt in this work is $h = 570$\,km. Fortunately, the uncertainty in this value only affects the physical scale inferred for the irregularities, and not the orientation or the magnitude of $\nabla\Sigma$. This can be seen directly from Equations (\ref{eq:trafo2a}) and (\ref{eq:trafo2b}), which are independent of $h$.

The value of $i$ can be set to be the magnetic inclination, which is known from geomagnetic reference models \citep[e.g.][]{AGRF2010}. However, this value can be directly fitted using MWA data without any prior knowledge of the magnetic field, serving as a useful quantitative cross-check of the assumption that the irregularities are field aligned. We did this by finding the $i$ that maximised the ``parallelness'' of features in the resulting $\nabla\Sigma$ distribution. In more quantitative terms, we made use of the power spectrum technique developed by \citet{Loi2015a_mn2e} applied to the new $\nabla\Sigma(\xi,\eta)$ vector field to obtain the total power of fluctuations with $k_\eta = 0$, where $k_\eta$ denotes spatial frequency along the $\eta$ direction. These are the fluctuation modes whose phase fronts are parallel to the $\eta$ axis. In the notation of \citet{Loi2015a_mn2e} (sections 2.3 and 2.4), but replacing $x \to \xi$, $y \to \eta$, $\Delta x \to \partial_\xi\Sigma$ and $\Delta y \to \partial_\eta\Sigma$, the optimisation problem can be stated quantitatively as the desire to find the $i$ maximising
%\begin{linenomath*}
\begin{equation}
  F(i) = \frac{V}{N_\xi N_t} \sum_{\ell=1}^{N_\xi} \sum_{n=1}^{N_t} \big[ \mathcal{P}_{\ell 1 n}(0^\circ) + \mathcal{P}_{\ell 1 n}(90^\circ) \big] \:. \label{eq:optimise_i}
\end{equation}
%\end{linenomath*}
This is the discrete version of the sum of the integrals of the power spectral density over $k_\xi$ and $\omega$ for the two scalar fields $\partial_\xi\Sigma$ and $\partial_\eta\Sigma$ (corresponding to the arguments of $0^\circ$ and $90^\circ$ in the above expression, respectively). The units of $\mathcal{P}$ in the current work are el$^2$\,m$^{-4}$\,s. We gridded the data at a resolution of 2\,km. 

The quantity $F$ is plotted versus $i$ in Figure \ref{fig:optimal_inc}. This exhibits a maximum near $i = (59 \pm 1)^\circ$. Although this seems to differ slightly from the zenith value of the magnetic inclination at a height of 570\,km, which is $60.3^\circ$ \citep{AGRF2010}, this can be explained by noticing that the north-south physical span of the plane within the FOV is in fact biased towards the north (by $\sim$100\,km, as can be seen in Figure \ref{fig:shading_xieta}). Since the magnetic inclination is shallower towards the north, the average inclination over the FOV should be slightly smaller than the zenith value: at 50 ($= 100 \cos 60^\circ$)\,km north of the MWA it is $59.7^\circ$, and incorporating in addition the curvature of the Earth away from an observer by $\sim 0.5^\circ$ every 50\,km, it is within expectation that the best fit value of $i$ lies closer to $59^\circ$ than $60^\circ$. In all subsequent analyses we adopted a value of $i = 59^\circ$. Our results change very little with variations of this value by $\sim 1^\circ$.

As a final comment in this section, we note that the near-constancy of the magnetic inclination along the line through zenith precludes the possibility of inferring $h$ through a fit to $i$. This has previously been done for similar structures (using other interferometers) when lines of sight have been suitably oblique \citep{Jacobson1996, Hoogeveen1997, Hoogeveen1997a}. However, the advantage here is that we have been able to quantitatively verify that the irregularities are indeed consistent with field alignment (to within $\sim 1^\circ$), whereas the previous works held this as a model assumption.

\begin{figure}
  \centering
  \includegraphics[width=0.5\textwidth]{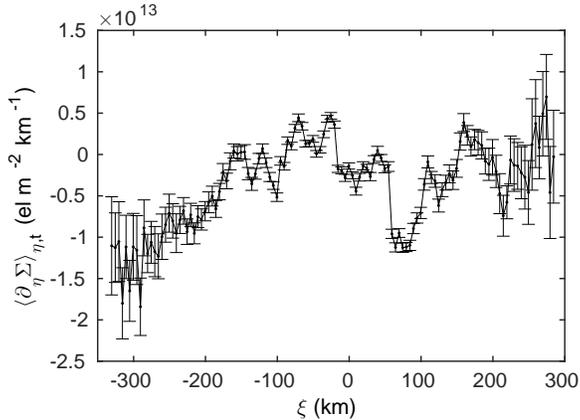}
  \caption{The average (over time and $\eta$) $\eta$-gradient, as a function of $\xi$. Physically, this is the field-aligned density gradient of each flux tube as a function of its longitudinal position. Points have been binned in 5-km intervals along the $\xi$-direction, and a minimum of 10 points in a bin is required for the bin to be shown on the plot. Error bars represent the standard error (standard deviation divided by the square root of the number of points) for each bin.}
  \label{fig:av_etagradient_vs_xi}
\end{figure}

\section{Super-MWA Structures}\label{sec:superMWA}
\subsection{Morphology}
Figure \ref{fig:shading_xieta} shows the spatial distributions of $\partial_\xi\Sigma$ and $\partial_\eta\Sigma$, the two orthogonal components of the $\nabla\Sigma$ vector field, for a representative snapshot from the \citet{Loi2015_mn2e} dataset. Movie S1 (see supplementary material online) shows the time evolution for the whole dataset. For the purposes of visualisation, the discretely sampled data have been interpolated onto a uniform 5\,km grid. This spacing is about half the average pierce-point separation. A representative distribution of the actual pierce points is shown in Figure \ref{fig:arrow}b, where one arrow corresponds to one pierce point. The sampling becomes sparser towards the north (higher altitudes) because the same solid angle subtends a greater physical area at a larger distance. 

The left and right panels correspond to the field-transverse (i.e.~direction in the plane perpendicular to the magnetic field) and field-aligned density gradients, respectively. We see that vertical (i.e.~$\eta$-parallel) striations are prominent in the field-transverse component of $\nabla\Sigma$, but not the field-aligned component. Typical length scales are about an order of magnitude smaller in the field-transverse than field-aligned directions. This is unsurprising because the Lorentz force acts perpendicular to the magnetic field lines and can only sustain against diffusion in this direction. Gradients along the $\eta$ direction would be caused by other effects, such as modulations by hydrodynamic or MHD waves \citep{Poole1988, Kazimirovsky2002, Menk2007, Fritts2008a, Waters2009}. This illustrates the usefulness of the $(\xi,\eta)$ basis in allowing us to decompose structures shaped by different physical processes. 

Also apparent from Figure \ref{fig:shading_xieta} is the difference in magnitudes of field-transverse and field-aligned gradients. Field-transverse gradients are about an order of magnitude steeper than field-aligned gradients, which is similarly consistent with physical expectations. In Movie S1 it can be seen that both components of $\nabla\Sigma$ increase in magnitude over time, indicating that the observation interval captured a period of growth. We discuss this further in Section \ref{sec:growth}.

\subsection{Field-aligned Density Gradients}
Under diffusive (i.e.~hydrostatic) equilibrium, one expects $\Sigma$ to decrease with altitude ($\partial_\eta\Sigma < 0$) along each flux tube. However, we can see from Figure \ref{fig:shading_xieta} and Movie S1 that $\partial_\eta\Sigma$ exhibits variations along $\eta$ on scales of several hundred kilometres, and that this can apparently take both positive and negative values (red and blue hues, respectively). However, we have measured angular offsets not with respect to an absolute reference but rather the time-averaged position, which subtracts away fluctuations that are stationary with respect to the celestial sphere. (The choice to use the time-averaged position as the reference, as has been done in all previous MWA works, stems from the inability to obtain an accurate absolute reference for this analysis; see \citet{Loi2015b_mn2e}, section 2.3 for details.) This implies that the $\partial_\eta\Sigma$ values are not absolute, but should rather be interpreted as fluctuations about some unmeasured global mean. 

Despite this limitation, there are still useful conclusions that can be drawn from the data. The rotation of the Earth causes each pierce point to drift roughly east-west (i.e.~in the $\xi$-direction) through the screen, and so a large fraction, particularly in the central $\xi$-range of the area scanned, eventually drift across (``sample'') each flux tube. Subtracting the mean thus retains fluctuations on scales smaller than the east-west extent of the FOV, implying that relative differences in $\partial_\eta\Sigma$ between flux tubes can be measured reliably.

We investigated the flux tube dependence of the field-aligned gradient by binning the data points gathered from all snapshots by $\xi$-value and then computing the average $\partial_\eta\Sigma$ value for each bin. The result is shown in Figure \ref{fig:av_etagradient_vs_xi}, with the characteristic spread within each bin represented by the error bars. There is evidence for systematic variations in $\partial_\eta\Sigma$ on east-west scales of order several tens of kilometres. This is comparable to the observed widths of the FAIs, and indicates that plasma scale heights can differ measurably between flux tubes that are several tens of kilometres apart. 

To explain the lack of diminished density contrast at high zenith angles, \citet{Loi2015_mn2e} argued that the thickness of the sheet could not be more than a factor of $\sim$3 larger than the spacing of the structures. If the thickness is $\sim$100\,km and the background electron density is $\sim 10^4$\,el\,cm$^{-3}$ (characteristic of the nighttime midlatitude topside ionosphere; \citet{Schunk2009}), then under the model assumption that all density variations are confined to the screen implies a relative variation in the scale height of $\sim$100\,km between flux tubes within the FOV. Given that absolute scale heights of several hundred kilometres have been measured at similar latitudes, altitudes and times of day \citep{Thomas1966, Watt1966}, this appears to be a substantial variation. However the actual thickness of the screen is largely uncertain, and a thinner screen would yield a proportionally smaller estimate. The physical origin for the local variations in scale height could be temperature differences between flux tubes, perhaps a result of variations in the conductivity tensor combined with storm-related field-aligned currents that produce uneven heating of the plasma, or longitudinally varying rates of impact ionisation. 

\begin{figure*}
  \centering
  \includegraphics[clip=true, trim=2cm 0cm 2cm 0cm, width=\textwidth]{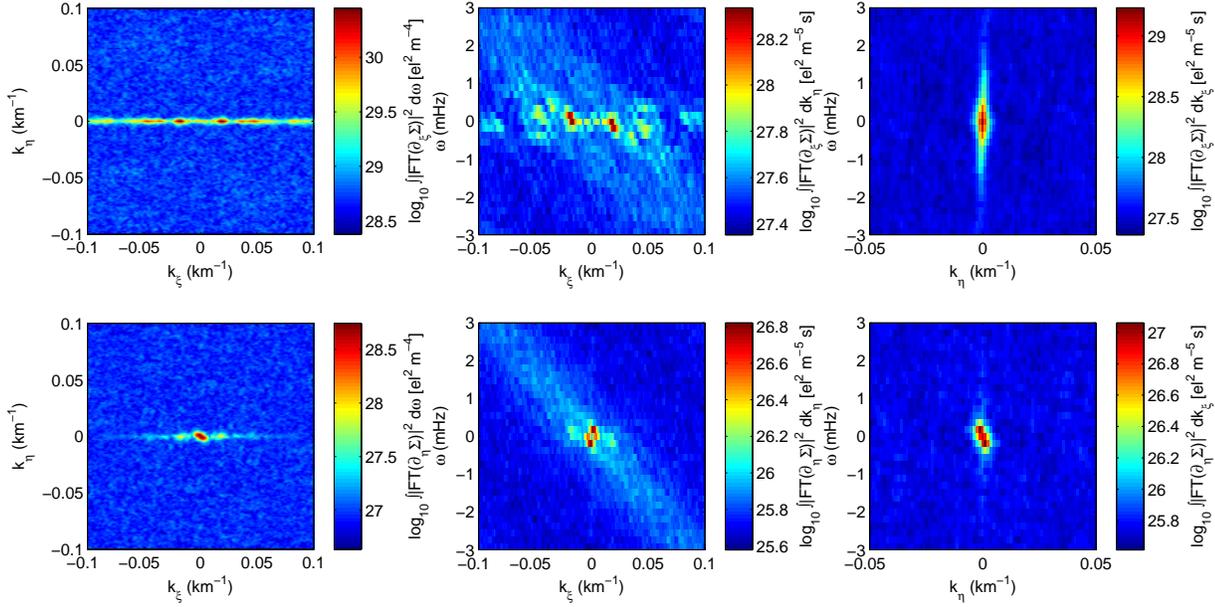}
  \caption{Logarithm of the power spectral density for the $\xi$-component (top row) and $\eta$-component (bottom row) of the density gradient field $\nabla\Sigma$, collapsed from three dimensions down to two by integrating out $\omega$ (left), $k_\eta$ (middle) and $k_\xi$ (right). Note that these have been zoomed in to the central regions; the absolute Nyquist frequencies are 0.25\,km$^{-1}$ for both the $k_\xi$ and $k_\eta$ axes and 4.2\,mHz for the $\omega$ axis. The corresponding response function is shown in Figure \ref{fig:powerspec_response}.}
  \label{fig:powerspec}
\end{figure*}

\begin{figure*}
  \includegraphics[clip=true, trim=2cm 0cm 2cm 0cm, width=\textwidth]{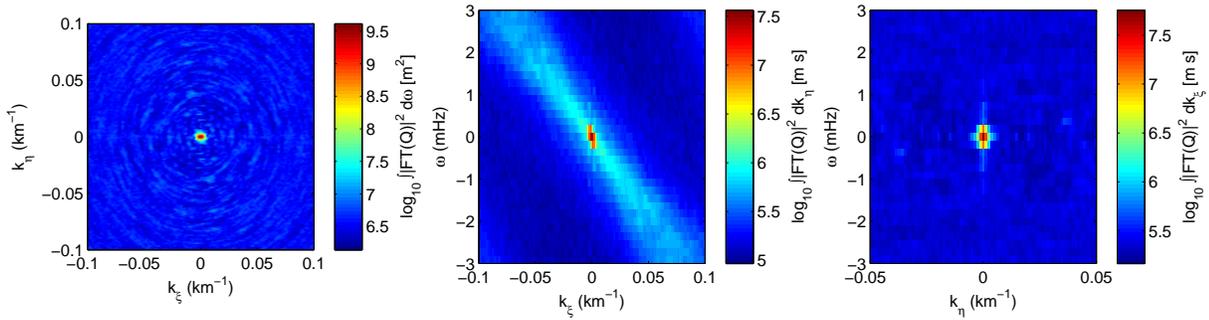}
  \caption{The response function for the power spectra shown in Figure \ref{fig:powerspec}. The three panels correspond to integrating out $\omega$ (left), $k_\eta$ (middle) and $k_\xi$ (right). The diagonal ringing feature in the middle panel results from the rotation of the Earth and consequent drift motion of measurement points through the screen (see \citet{Loi2015a_mn2e} for a more detailed explanation of its origin).}
  \label{fig:powerspec_response}
\end{figure*}

\subsection{Power Spectra and Wavelike Behaviour}\label{sec:powerspec}
This and the following subsection present quantitative measurements of the fluctuation properties (wavelengths, periods and phase velocities) of the $\nabla\Sigma$ vector field using the power spectrum technique developed by \citet{Loi2015a_mn2e}. Spatial coordinates $x$ and $y$ are replaced by $\xi$ and $\eta$, and the transformed quantity is $\nabla\Sigma$ (rather than $\nabla_\perp$TEC). We chose a grid size of 2\,km, which given a mean sampling density of approximately one pierce point per 100\,km$^2$ on the screen, corresponds to spatially oversampling by a factor of about two. We computed two power spectra, one for $\partial_\xi\Sigma$ and another for $\partial_\eta\Sigma$. Each is a three-dimensional data cube with two axes for the $\xi$ and $\eta$ spatial frequencies (denoted by $k_\xi$ and $k_\eta$) and one axis for the temporal frequency $\omega$. Figure \ref{fig:powerspec} shows the reduced (from 3D to 2D) power spectral density distributions obtained by integrating along each of the axes in turn.

The response function for these power spectra, visualised in an identical manner, is shown in Figure \ref{fig:powerspec_response}. Its compactness assures us that the most of the structure seen in Figure \ref{fig:powerspec}, with the exception of the diffuse diagonal band running from top-left to bottom-right in the middle column, is genuine. The band is a typical feature in MWA power spectra and we refer the reader to \citet{Loi2015a_mn2e}, section 3.1 for details of its origin.

It can be seen that the greatest concentration of power is in the cells for which $k_\eta = 0$ (left column of Figure \ref{fig:powerspec}). This tells us that fluctuations in $\partial_\xi\Sigma$ take the form of corrugations varying almost purely along the $\xi$-direction, which is consistent with Figure \ref{fig:shading_xieta}. As mentioned earlier in Section \ref{sec:inclination}, the value of $i$ was chosen according to Equation \ref{eq:optimise_i}. This corresponds to choosing the $i$ that concentrates the most signal power into the $k_\eta = 0$ cells. Otherwise put, it is the act of choosing the basis in which as much information is captured in as small a subset of basis vectors as possible. The fact that the value of $i$ which achieves this coincides with the magnetic inclination provides the physical interpretation for the structures as FAIs. A separate subsection (Section \ref{sec:modes}) is devoted to an analysis of the structure within the $k_\eta = 0$ row of cells.

Comparing the top and bottom rows of Figure \ref{fig:powerspec} reveals that peaks tend to be 1--2 orders of magnitude more intense for $\partial_\xi\Sigma$ than $\partial_\eta\Sigma$. It can also be seen from the left and middle columns that the power spectral density for $\partial_\eta\Sigma$ is confined to smaller $k_\xi$ values (larger east-west scales) than $\partial_\xi\Sigma$. This is consistent with Figure \ref{fig:shading_xieta} and Movie S1.

Phase velocities can be measured from the slopes of features in the $\omega$-$k_\xi$ and $\omega$-$k_\eta$ planes (middle and right columns). A north-eastward propagating fluctuation in $\partial_\eta\Sigma$ can be identified, with a wavelength of $\sim$700\,km and a phase speed of about 200\,m\,s$^{-1}$. This may be an AGW whose presence is revealed only in the $\partial_\eta\Sigma$ component because this component is not overwhelmed by the FAI signal. In contrast, the modes with smaller wavelengths seen in $\partial_\xi\Sigma$ (top row, middle column) are associated with smaller slopes and therefore phase speeds. If one inspects the slopes of lines joining pairs of conjugate peaks one sees that there is a small scatter about $\omega/k_\xi = 0$\,m\,s$^{-1}$, with some slopes being positive and others negative. This suggests that different modes move with slightly different speeds (typical magnitudes are several metres per second), some drifting east and others west. Since the associated FAIs are likely to be \textit{in situ} stationary structures, this suggests the presence of east-west shear in the plasma at a rate of $\sim$10\,m\,s$^{-1}$, which could be a potential source of free energy for growth. 

\begin{figure*}
  \centering
  \includegraphics[width=0.9\textwidth]{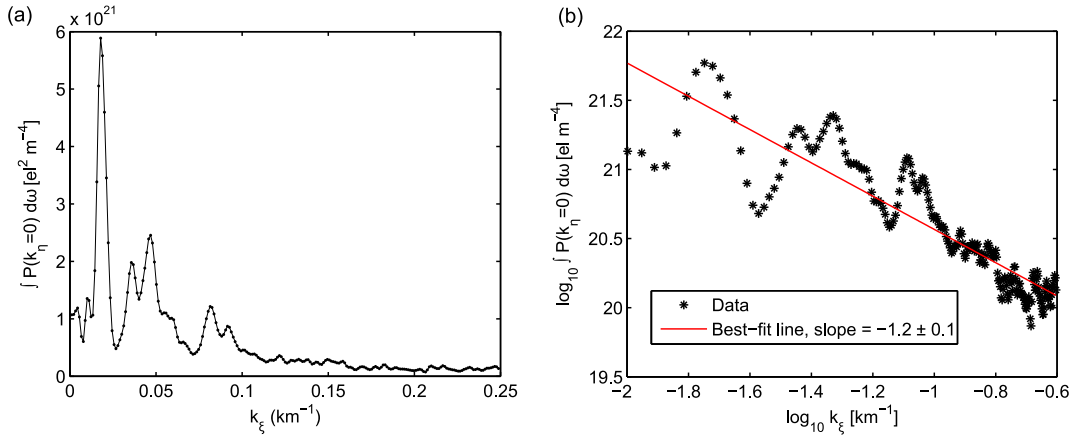}
  \caption{The power spectral density in the $k_\eta = 0$ row of cells from the top-left panel of Figure \ref{fig:optimal_inc}, plotted as a function of $k_\xi$ on (a) linear and (b) logarithmic axes. Since the power spectrum is inversion-symmetric, it is only necessary to display one half of the $k_\xi$ axis (we have chosen to plot positive $k_\xi$). Strong peaks at discrete spatial frequencies can be seen, indicating the presence of distinct modes. A fit to the power spectrum, corresponding to the red line in panel (b), suggests that it can globally be described by a power law of index $-1.2 \pm 0.1$.}
  \label{fig:keta0pow_vs_kxi}
\end{figure*}

\begin{figure}
  \centering
  \includegraphics[width=0.5\textwidth]{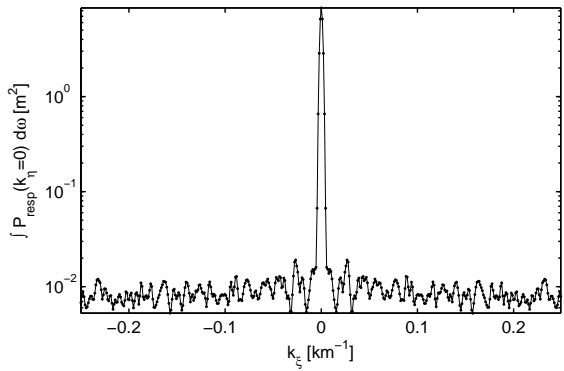}
  \caption{The response function corresponding to the one-dimensional power spectrum shown in Figure \ref{fig:keta0pow_vs_kxi}. Noisy fluctuations have been made visible by the use of a logarithmic scale on the vertical axis. The central peak is about three orders of magnitude above the noise and has a width of 3--4 frequency increments.}
  \label{fig:keta0pow_response}
\end{figure}

\subsection{Spatial Periodicities}\label{sec:modes}
The concentration of power into the modes with $k_\eta = 0$ through use of the $(\xi,\eta)$ basis allows us to isolate this region of the $\partial_\xi\Sigma$ power spectrum for further analysis. The $\omega$-integrated power spectral density distribution as a function of $k_\xi$ for the $k_\eta = 0$ modes is shown in Figure \ref{fig:keta0pow_vs_kxi}. One can identify prominent peaks at distinct spatial frequencies, whose heights roughly decrease with increasing wavenumber. The corresponding response function, shown in Figure \ref{fig:keta0pow_response}, is highly compact and would not be responsible for generating this complex structure.

The most prominent peak is at $k_\xi = 0.018$\,km$^{-1}$, corresponding to a wavelength of 56\,km (this periodicity can be visually identified in Figure \ref{fig:shading_xieta}). Additional peaks appear near $k_\xi$ = 0.036, 0.047, 0.081 and possibly 0.092\,km$^{-1}$, corresponding to wavelengths of 28, 21, 12 and 11\,km respectively. Note that although the wavenumber itself can be measured to a precision of 10$^{-3}$\,km$^{-1}$ (set by the spatial extent of the observed region), the overall error in these values is dominated by the systematic error arising from the uncertainty in $h$. This is large, of order 10\%, and enters through the role of $h$ in setting the linear scale of the $(\xi,\eta)$ coordinates with respect to the data, which can be seen from Equation (\ref{eq:trafo1}). 

Although the $k_\eta = 0$ region of the $\partial_\xi\Sigma$ power spectrum has a complicated structure containing multiple peaks, the envelope appears to decay with increasing $k_\xi$. If we assume a power-law functional form, the best-fit index turns out to be $-1.2 \pm 0.1$. Note that this applies to the power spectrum of the $\xi$-gradient of $\Sigma$. Because the Fourier transform of the $\xi$-gradient of a function is $-ik_\xi$ times the Fourier transform of the function itself, the power spectrum of $\partial_\xi\Sigma$ is $k_\xi^2$ times the power spectrum of $\Sigma$ and so the corresponding index of the \textit{density} power spectrum is $-3.2$. This may be consistent with the 3D Kolmogorov index of $-11/3$ \citep{Kolmogorov1941} to within experimental uncertainty if one acknowledges the possibility of additional sources of error such as spectral leakage, which tends to systematically flatten the spectrum (see \citet{Loi2015a_mn2e}, Appendix A). Moreover, it is uncertain as to where in wavenumber space a turbulent cascade might begin. It may be that the peaks seen at $k_\xi < 0.1$\,km$^{-1}$ reflect unstable modes where energy is being injected, and that the cascade only develops at higher wavenumbers. Indeed, the region where $k_\xi > 0.1$\,km$^{-1}$ appears to possess a slightly steeper index of $-1.4 \pm 0.2$.

Given the evidence for velocity shear (Section \ref{sec:powerspec}), it is possible that the structures have grown through a Kelvin-Helmholtz instability. The dominant wavelength in this case is around eight times the width of the shear layer, so if the primary mode is the one at $\sim$60\,km then one infers a thickness for the shearing region of order 10\,km. The origin of this shear may be in the neutral atmosphere, or alternatively, the periodicities may reflect those of seed disturbances (e.g.~AGWs), where the associated density and conductivity perturbations have mapped upwards along field lines to generate the structures observed.

\begin{figure}
  \centering
  \includegraphics[width=0.5\textwidth]{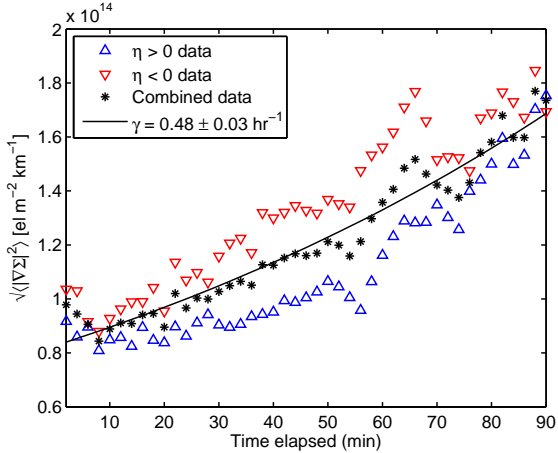}
  \caption{The RMS of density gradients measured through source position offsets, and therefore indicating the characteristic amplitude of fluctuations on scales exceeding the size of the MWA ($\sim$1\,km), as a function of time. Black asterisks correspond to the RMS value calculated over the whole FOV, whereas the upwards-pointing (blue) and downwards-pointing (red) triangles correspond to restrictions to the regions northward ($\eta > 0$) and southward ($\eta < 0$) of zenith, respectively. The slope of the plane towards the north means that this also reflects a split in altitude, with blue triangles being for higher altitudes. The black solid line is an exponential fit ($\propto \mathrm{e}^{\gamma t}$) through the asterisks. The best-fit value for the growth rate $\gamma$ corresponds to a timescale of about 2\,hr.}
  \label{fig:superMWA_totgrad_vs_time}
\end{figure}

\subsection{Growth over Observing Interval}\label{sec:growth}
As previously noted, the density gradients associated with the FAIs grow over time. This can be quantified by plotting the characteristic $\nabla\Sigma$ value over the FOV as a function of time, shown using black asterisks in Figure \ref{fig:superMWA_totgrad_vs_time}. To enable a quantitative comparison with the small-scale gradients defined in Equation (\ref{eq:broadwid}) (see later in Section \ref{sec:broadening}), it is the RMS value of $\nabla\Sigma$ that has been plotted. We obtained the characteristic growth rate $\gamma$ by fitting to the data an exponential curve (black solid line) of the form $\propto \mathrm{e}^{\gamma t}$, where $t$ is time. The best-fit growth rate is $\gamma = 0.48 \pm 0.03$\,hr$^{-1}$, corresponding to a growth timescale of about 2\,hr. 

We also examined the separate trends for high ($\eta > 0$) and low ($\eta < 0$) altitudes, plotted as the blue and red triangles in Figure \ref{fig:superMWA_totgrad_vs_time}. While growth on a similar timescale is evident in both regimes, the two curves are noticeably offset with characteristic gradients being about 20\% larger at lower altitudes. This could arise if each flux tube were associated with a fixed-percentage fluctuation of the background density; a global decrease in background density with altitude would then explain the separation between the curves.

\section{Sub-MWA Structures}\label{sec:subMWA}

\subsection{Scintillation}\label{sec:scintillation}
Because an interferometer combines signals coherently, it is susceptible to destructive interference and decorrelation if the phase variation of an incoming signal over the array deviates significantly (by $>$1\,radian) from a linear ramp. This effect is similar to the constructive and destructive interference that results if ray paths are perturbed during propagation to the point where they cross before reaching the detector (cf.~Fraunhofer diffraction). The critical size scale of irregularities on which $\sim$1\,rad associated phase perturbations cause this to occur (i.e.~the Fresnel scale) is around 400\,m for the parameters relevant here. These effects give rise to scintillation, where the apparently brightness of the source fluctuates erratically by a significant fraction of the true brightness. The timescale of this fluctuation is given by the spatial scale of the irregularities divided by the drift speed of the sight lines.

\begin{figure*}
  \centering
  \includegraphics[width=0.9\textwidth]{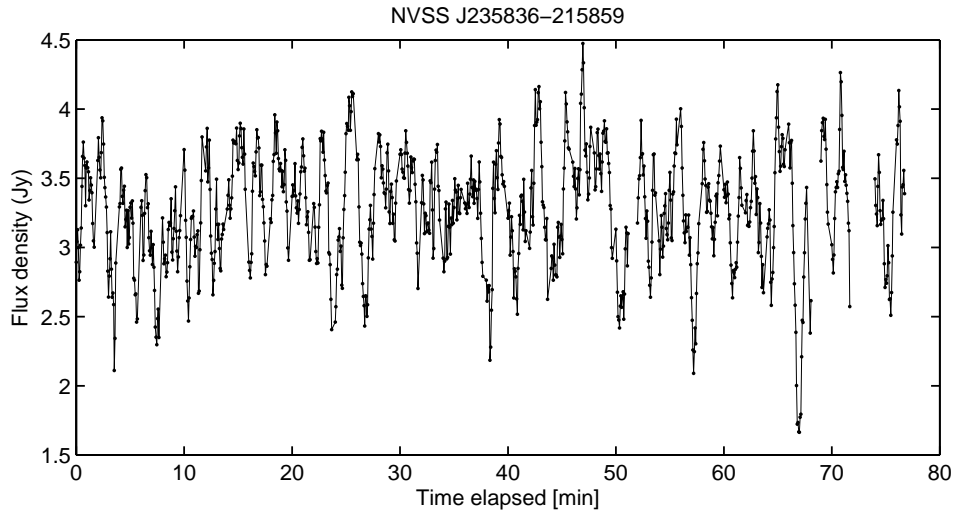}
  \caption{A representative radio light curve (flux density as a function of time) of a randomly-selected source, exhibiting a modulation index of order unity on short (tens of seconds) timescales. This is characteristic of diffractive scintillation, evidencing significant structure on sub-array scales. Note that this plot was obtained using images formed at a cadence of 4\,s. Similar rapidly-varying light curves are observed for all other sources in the dataset. The error on each flux measurement, given by the local pixel-to-pixel RMS noise, is 0.2\,Jy.}
  \label{fig:lightcurve_example}
\end{figure*}

Figure \ref{fig:lightcurve_example} shows the radio light curve for a moderately bright (S/N ratio of 15) source in the data. Large modulations, of up to 40\% the mean brightness of the source and significantly larger than the uncertainty of the measurement ($\sim$0.2\,Jy, given by the local pixel-to-pixel noise), can be observed to occur on tens of second timescales. Rapid brightness variations are similarly seen for many other sources in the field, and can be interpreted as evidence for substantial phase structure on sub-MWA scales. Although it is in principle possible to attempt a temporal power spectrum analysis of these light curves, this is problematic for several reasons. These include the presence of 8-s data gaps for every 2-min block, abrupt beamformer changes at 30-min intervals, and systematic uncertainties in the direction-dependent instrument gain which contaminate the power spectrum across a wide range of frequencies. Suffice to say that no obvious peaks appear in the temporal power spectra (not shown) that might point to a particular diffractive scale. It is possible that this varies over the FOV, but not much more can be said in this regard.

\begin{figure*}
  \centering
  \includegraphics[width=0.9\textwidth]{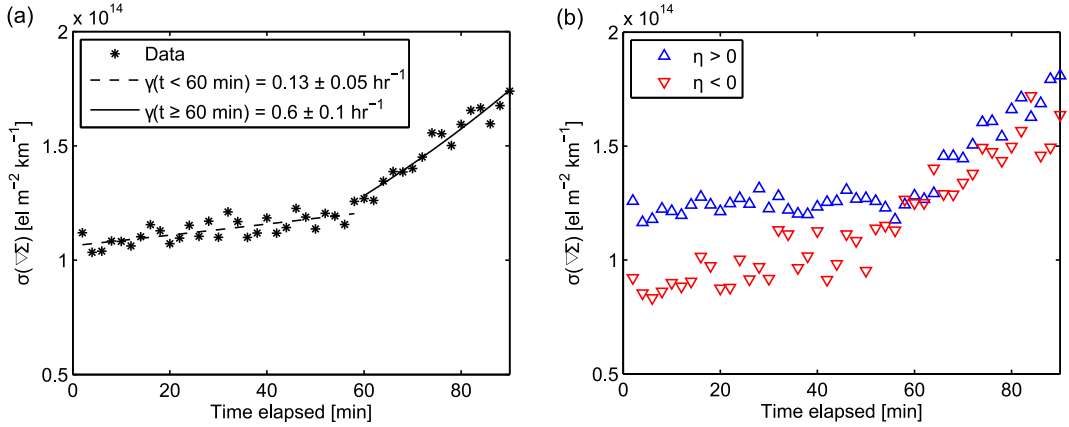}
  \caption{The RMS of density gradients measured through the angular broadening of otherwise unresolved sources, and therefore on scales below the size of the MWA ($\sim$1\,km), as a function of time. Panel (a) shows the trend for sources over the whole field of view, whereas (b) shows the trend for two sets of sources separated according to whether they lie north (blue) or south (red) of zenith. There appears to be an increase in growth rate near time $t = 60$\,min in panel (a), which coincides with the time when the two growth curves in panel (b), initially separate, converge. To quantify this apparent change in growth rate, the data before and after $t = 60$\,min have been fitted separately with exponential functions ($\propto \mathrm{e}^{\gamma t}$). These are the dashed and solid lines in (a), with corresponding best-fit growth rates $\gamma$ indicated in the figure legend.}
  \label{fig:subMWA_totgrad_vs_time}
\end{figure*}

\subsection{Angular Broadening}\label{sec:broadening}
The quantity $\sigma(\nabla\Sigma)$, defined in Equation (\ref{eq:broadwid}) and representing the characteristic spread in density gradients for FAIs on sub-MWA scales, is plotted as a function of time in Figure \ref{fig:subMWA_totgrad_vs_time}. We observe that this increases with time, and at a similar average rate of growth as for the super-MWA structures. However, the growth curve appears (at least visually) to be segmented into two portions associated with different growth rates, the transition occurring near the 60-min mark. We have chosen to fit these two portions separately, and find that the growth rate (given by the slope of the trend) is consistent with undergoing an increase at $t = 60$\,min by a factor of 4--5. The corresponding timescales of growth are 7--8\,hr before $t = 60$\,min, subsequently dropping to 1--2\,hr.

Splitting the data into high ($\eta > 0$) and low ($\eta < 0$) altitude regions reveals another property of the apparent transition at $t = 60$\,min. The two growth curves are initially separate, with $\sigma(\nabla\Sigma)$ larger at higher altitudes, but then converge at the transition point (see Figure \ref{fig:subMWA_totgrad_vs_time}b). The initial difference in $\sigma(\nabla\Sigma)$ between the two regimes could be explained by a larger dissipation rate at lower altitudes, where the environment is more collisional. The subsequent convergence of the two curves might then be attributed to the increase in growth rate, with the new growth rate being much larger than the difference in dissipation rates between high and low altitudes. Although the situation appears to be reversed compared to super-MWA structures in that higher altitudes are associated with larger gradients, this could be explained by the greater importance of dissipative effects on smaller scales. 

It is possible that these small-scale structures derived their energy for formation from the same source as for the larger FAIs. However, large-scale FAIs with density gradients comparable to or greater than this dataset have been previously observed by the MWA that were not accompanied by scintillation, which is uncommon at these observing frequencies \citep{Loi2015b_mn2e, Loi2016}. A number of studies that simultaneously monitored the occurrence of VLF whistlers and VHF scintillation have found the two to be uncorrelated, which suggests that large- and small-scale structures may not share the same physical origin \citep{Singh1994, Patel2010}. It may be that the simultaneous presence of both large- and small-scale irregularities in this dataset is coincidental. However, the observation here that the two share similar growth trends supports the idea that they share a common energy source, so perhaps the small-scale irregularities belong to the high-frequency end of the cascade seen on larger scales (Section \ref{sec:modes}). Ascertaining the conditions necessary for simultaneous or independent formation of FAIs of different sizes will require a broader study that is beyond the scope of this work.

It is curious to notice that the value of $\sigma(\nabla\Sigma)$, measuring the power on sub-MWA scales, is very similar to the RMS density gradient for super-MWA scales (compare the vertical axes of Figures \ref{fig:superMWA_totgrad_vs_time} and \ref{fig:subMWA_totgrad_vs_time}), which may be a coincidence. If we can assume that the power-law index of $-1.2$ measured for large-scale density gradient fluctuations can be extrapolated down to small scales (i.e.~that these are part of the same spectrum), this gives us a way of estimating the inner scale. Taking the wavenumbers associated with the outer scale and MWA to be $k_\text{out} = 10^{-4}$\,m$^{-1}$ and $k_\text{MWA} = 10^{-3}$\,m$^{-1}$, and equating the total power (proportional to $\int k^{-1.2} \,\mathrm{d}k$) on super- and sub-MWA scales, we obtain $k_\text{in} \sim 0.1$\,m$^{-1}$ for the inner scale. The corresponding length is $\sim$10\,m, comparable to ion gyroradii.

\section{Summary and Outlook}\label{sec:conclusion}
We have introduced a new approach for studying FAIs that transforms interferometric TEC gradient data into a physically meaning reference frame, corresponding to the magnetic shell tangent plane. Applying the transformation to the 15 October 2013 dataset of \citet{Loi2015_mn2e}, we demonstrated its utility for decomposing behaviour along and across magnetic field lines through the construction of a suitable set of coordinates $(\xi,\eta)$, where the $\xi$ and $\eta$ directions are perpendicular and parallel to the field, respectively. We also devised a method for computing the best value for the inclination of the plane from the data, and verified that it is in good agreement with the magnetic inclination. Through our analyses we found that:
\begin{itemize}
  \item The transformation to $(\xi,\eta)$ confirms with much greater precision than previous work that the striations seen by the MWA on 15 October 2013 were oriented along the geomagnetic field;
  \item $\xi$-gradients were around an order of magnitude steeper than $\eta$-gradients and showed structuring on spatial scales of several tens of kilometres, while $\eta$-gradients were structured on much larger scales of several hundred kilometres;
  \item The plasma scale height, quantified through the $\eta$-gradients, varied by up to $\sim$100\,km between flux tubes within the FOV, possibly reflecting spatial variations in plasma temperature;
  \item A differential drift of $\sim$10\,m\,s$^{-1}$ in the east-west direction could be identified through power spectrum analysis, possibly providing the energy for growth of the FAIs through a shear-driven instability;
  \item The spatial power spectrum showed multiple periodicities on scales between 10--60\,km, which may correspond to those of seed fluctuations (e.g.~atmospheric waves) or perhaps the favoured wavelengths of a plasma instability;
  \item Fitting the power spectrum of $\xi$-gradients with a power law yielded an index of $-1.2 \pm 0.1$, slightly flatter than the Kolmogorov value of $-5/3$;
  \item Growth of the FAIs on both large and small spatial scales was observed over the interval, and we measured the characteristic timescales to be several hours;
  \item Many background radio sources exhibited strong scintillations, implying significant structure on small (sub-array) scales;
  \item Angular broadening analyses revealed small-scale irregularities to be highly anisotropic and aligned along the magnetic field;
  \item Comparable amounts of fluctuation power were present on large and small scales (relative to the size of the array), allowing us to estimate an inner scale of $\sim$10\,m by extrapolating the power law obtained for large-scale structure;
  \item The strengths of density gradients exhibited an altitude dependence, being greater at lower altitudes for large-scale structure and greater at higher altitudes for small-scale structure, which might be explained through a size-dependent importance of dissipative effects.
\end{itemize}

The main weakness of our model is the assumption that irregularities are confined to a thin layer. This may be a fair approximation for the interval analysed here, but there may be periods in which irregularities reside on multiple layers, a situation that would escape detection by the MWA. Because of its small size, the MWA can only perform a crude height localisation through parallax analysis to obtain a single characteristic value for the altitude \citep{Loi2015_mn2e}. Interferometers that are more spatially extended, such as the Low Frequency Array \citep{vanHaarlem2013} and the future Square Kilometre Array (SKA) \citep{Dewdney2009}, would be capable of conducting a more detailed three-dimensional mapping of the ionosphere through e.g.~tomographic inversion \citep{GaussiranII2004, Koopmans2010}.

A convenient property of our model is that it is self-contained: the two input parameters (height and inclination) can be fitted using the data alone. The concepts presented here may pave the way for improvements to ionospheric calibration procedures for next-generation radio telescopes. The two-parameter inclined screen model enables most of the FAI fluctuation power to be captured into unidirectional plane-wave modes aligned with the Earth's magnetic field. Such a model may be a viable starting point for the development of compact calibration strategies for the SKA, since a relatively small number of basis elements would be needed to represent FAI-related distortions in the data. The ability of interferometers like the MWA to probe ionospheric density structures at great breadth and detail underscores them as rich sources of geophysical information; we hope for their continued application to this end in time to come.

%%% End of body of article:

%%%%%%%%%%%%%%%%%%%%%%%%%%%%%%%%
%% Optional Appendix goes here
%
% \appendix resets counters and redefines section heads
% but doesn't print anything.
% After typing \appendix
%
%\section{Here Is Appendix Title}
% will show
% Appendix A: Here Is Appendix Title
%

\appendix

\section{Transformation equations}\label{sec:trafo}
Here we present the mathematical details of the transformations applied to the data, namely the coordinate conversion from astronomical $(\alpha,\delta)$ coordinates to the Earth-based $(\beta,\gamma)$ coordinates and then the mapping onto the $(\xi,\eta)$ plane. Since the latter two systems are non-standard (to our knowledge), their descriptions have been included here for the sake of reproducibility of our work.

The coordinates $(\alpha,\delta)$ form a spherical coordinate system whose pole is aligned with the rotation axis of the Earth, $\alpha$ and $\delta$ being longitude and latitude, respectively. Zero longitude is defined to pass through the vernal equinox, a point fixed with respect to the stars. Like $(\alpha,\delta)$ coordinates, $(\beta,\gamma)$ coordinates also form a spherical system, with $\beta$ and $\gamma$ being longitude and latitude respectively. However, the pole of the $(\beta,\gamma)$ system is aligned with geographic east-west, and zero longitude is defined to pass through the zenith. These coordinates are therefore observer-dependent, differing from the more familiar Az/El coordinates by a $90^\circ$ rotation. They are illustrated in Figure \ref{fig:geometry}.

Transforming between $(\alpha,\delta)$ and $(\beta,\gamma)$ involves a simple rotation operation. To achieve this, one must know two things: the local sidereal time (LST) of the observation, and the geographic latitude $\Lambda$ of the observing location. Defining $\alpha' \equiv \alpha - \mathrm{LST}$, one then computes $\mathbf{x}' = \Phi \mathbf{x}$ where
%\begin{linenomath*}
\begin{equation}
  \Phi = \begin{pmatrix}
    \cos\Lambda & 0 & \sin\Lambda \\
    -\sin\Lambda & 0 & \cos\Lambda \\
    0 & -1 & 0
  \end{pmatrix} ,\; 
  \mathbf{x} = \begin{pmatrix}\cos\delta \cos\alpha' \\
    \cos\delta \sin\alpha' \\ 
    \sin\delta
  \end{pmatrix}
  \label{eq:trafo0}
\end{equation}
%\end{linenomath*}
and inverts $\mathbf{x}' = (\cos\gamma \cos\beta, \cos\gamma \sin\beta, \sin\gamma)$ for $\beta$ and $\gamma$. At the MWA site, $\Lambda = -26.7^\circ$. One can see that the transformation operation is time-dependent, reflecting the fact that the two coordinate systems move with respect to one another (since the Earth rotates with respect to the stars).

Given that the conversion from $(\alpha,\delta)$ to $(\beta,\gamma)$ is nothing more than a basis transformation, the $(\beta,\gamma)$ representation of a $\nabla_\perp$TEC vector given its $(\alpha,\gamma)$ representation can be obtained by applying the above operations to the endpoints of the corresponding angular offset vectors. That is, if a source whose true position lies at $(\alpha_0,\delta_0)$ is observed to be displaced by $(\Delta\alpha,\Delta\delta)$, then we would use the mapping described above to take $(\alpha_0,\delta_0) \to (\beta_0,\gamma_0)$ and $(\alpha_0+\Delta\alpha,\delta_0+\Delta\delta) \to (\beta_1, \gamma_1)$. Then $\nabla_\perp\mathrm{TEC} = (\partial_\beta\mathrm{TEC},\partial_\gamma\mathrm{TEC})$ can be obtained by substituting $\Delta\theta = ([\beta_1-\beta_0]\cos\gamma_0, \gamma_1-\gamma_0)$ into Equation (\ref{eq:displacement}).

Next, we specify how to obtain $\xi$, $\eta$, $\partial_\xi\Sigma$ and $\partial_\eta\Sigma$ from $\beta$, $\gamma$, $\partial_\beta$TEC and $\partial_\gamma$TEC. Unlike the transformation from $(\alpha,\delta)$ to $(\beta,\gamma)$, this is not merely a change of basis but a mapping between different vectors in 3D space. The geometrical setup is illustrated in Figure \ref{fig:geometry}. Once $h$ and $i$ are chosen, $(\xi,\eta)$ can be found by projecting $(\beta,\gamma)$ onto the screen:
%\begin{linenomath*}
\begin{equation}
  \begin{pmatrix}
    \xi \\
    \eta
  \end{pmatrix} = \frac{h}{\cos(\beta+i)}
  \begin{pmatrix}
    \cos i \tan\gamma \\
    \sin\beta
  \end{pmatrix} \:.
  \label{eq:trafo1}
\end{equation}
%\end{linenomath*}

Finding $\partial_\xi\Sigma$ and $\partial_\eta\Sigma$ requires a more involved sequence of operations. It is \textit{not} correct to simply project the endpoints of an angular displacement vector onto the screen and take the resulting separation to represent $\nabla\Sigma$, because Equation (\ref{eq:displacement}) no longer applies. To understand how $\nabla_\perp$TEC should map to $\nabla\Sigma$, observe that the local gradient vector of any scalar field defined on a surface measured from a certain vantage point is inversely proportional to the apparent distance between two contours (i.e.~lines joining points of equal value). Since the spacing between two given contours on a surface is foreshortened by the cosine of the angle at which the surface is viewed, the corresponding gradient vector will be \textit{forelengthened} by this factor. If we denote the normal vector to the surface by $\hat{\mathbf{n}}$, this means that $\nabla_\perp$TEC must be elongated by a factor $|\hat{\mathbf{n}}\cdot\mathbf{x}'|$ along the viewing direction. For our simple 2D plane model, $\hat{\mathbf{n}} = (-\cos i, \sin i, 0)$. We obtained (after some algebra) the following expressions for the components of $\nabla\Sigma$:
%\begin{linenomath*}
\begin{align}
  \partial_\xi \Sigma &= \sin\gamma \cos\gamma \sin(\beta+i) \cos(\beta+i) \nonumber \\
  & \times [\cos i \cos\gamma \cos(\beta+i) + \sin i - 1] \partial_\beta \mathrm{TEC} \nonumber \\
  & + \left\{\sin^2\gamma \cos\gamma \cos^2(\beta+i) [\cos i \cos(\beta+i) + \sin i] \right. \nonumber \\
  & + \left. \cos\gamma \sin^2(\beta+i)\right\} \partial_\gamma \mathrm{TEC} \label{eq:trafo2a} \\
  \partial_\eta \Sigma &= \cos(\beta+i) \left\{\sin^2(\beta+i) \cos^2\gamma \right. \nonumber \\
  & \times \left. [\cos(\beta+i) \cos\gamma \cos i + \sin i] + \sin^2\gamma\right\} \partial_\beta \mathrm{TEC} \nonumber \\
  & + \sin(\beta+i) \sin\gamma \left\{\cos^2(\beta+i) \cos^2\gamma \right. \nonumber \\
  & \times \left. [\cos(\beta+i) \cos i \cos\gamma + \sin i] - 1\right\} \partial_\gamma \mathrm{TEC} \:. \label{eq:trafo2b}
\end{align}
%\end{linenomath*}

Equations (\ref{eq:trafo2a}) and (\ref{eq:trafo2b}) are the equations used in the current work. For a more sophisticated model that deals with curvature effects, such as a dipole shell model, it is not necessarily the case that convenient closed-form expressions can be found. In the case of a dipole, $\hat{\mathbf{n}}$ is a function of $\xi$, and obtaining $\xi$ itself requires as an intermediate step solving a quintic in the radius of the dipole shell at the pierce point. This would be possible in practice through numerical means. However, the planar model appeals at the present stage because of its simplicity and compactness of implementation. Alternative or more refined models could be considered in future extensions to this work.

A useful generalisation of the planar model would be to allow for a non-zero magnetic declination $D$ (as implemented, the equations in fact align $\hat{\boldsymbol{\eta}}$ with geographic east-west). This could be achieved by introducing an alternative definition for the $(\beta,\gamma)$ coordinate system, where the pole is aligned with geomagnetic rather than geographic east-west. In this situation, $\Phi$ would be replaced by the more general matrix
%\begin{linenomath*}
\begin{equation}
  \Phi' = \begin{pmatrix}
    \cos\Lambda & 0 & \sin\Lambda \\
    -\sin\Lambda \sin D' & \cos D' & \cos\Lambda \sin D' \\
    -\sin\Lambda \cos D' & -\sin D' & \cos\Lambda \cos D'
  \end{pmatrix} \:,
  \label{eq:trafo0_D}
\end{equation}
%\end{linenomath*}
where $D' \equiv 90^\circ - D$. See that this reduces to $\Phi$ in Equation (\ref{eq:trafo0}) when $D = 0^\circ$. At the location of the MWA, $D$ is around $-0.2^\circ$ \citep{AGRF2010} and so has been neglected. However, similar analyses of data taken at locations where the magnetic declination is substantial (in excess of a few degrees) should use Equation (\ref{eq:trafo0_D}) rather than Equation (\ref{eq:trafo0}) if the form of Equations (\ref{eq:trafo1})--(\ref{eq:trafo2b}) is to be preserved.

%%%%%%%%%%%%%%%%%%%%%%%%%%%%%%%%%%%%%%%%%%%%%%%%%%%%%%%%%%%%%%%%
%
% Optional Glossary or Notation section, goes here
%
%%%%%%%%%%%%%%
% Glossary is only allowed in Reviews of Geophysics
% \section*{Glossary}
% \paragraph{Term}
% Term Definition here
%
%%%%%%%%%%%%%%
% Notation -- End each entry with a period.
% \begin{notation}
% Term & definition.\\
% Second term & second definition.\\
% \end{notation}
%%%%%%%%%%%%%%%%%%%%%%%%%%%%%%%%%%%%%%%%%%%%%%%%%%%%%%%%%%%%%%%%
%
%  ACKNOWLEDGMENTS

\begin{acknowledgments}
The MWA data supporting this paper are available on request submitted via email to the corresponding author at stl36@cam.ac.uk. CMT is supported under Australian Research Council's Discovery Early Career Researcher funding scheme (project number DE140100316) and the Centre for All-sky Astrophysics (an Australian Research Council Centre of Excellence funded by grant CE110001020). This scientific work makes use of the Murchison Radio-astronomy Observatory, operated by CSIRO. We acknowledge the Wajarri Yamatji people as the traditional owners of the Observatory site.  Support for the operation of the MWA is provided by the Australian Government Department of Industry and Science and Department of Education (National Collaborative Research Infrastructure Strategy: NCRIS), under a contract to Curtin University administered by Astronomy Australia Limited. We acknowledge the iVEC Petabyte Data Store and the Initiative in Innovative Computing and the CUDA Center for Excellence sponsored by NVIDIA at Harvard University.
\end{acknowledgments}

\end{article}
%
%
%% Enter Figures and Tables here:
%
% DO NOT USE \psfrag or \subfigure commands.
%
% Figure captions go below the figure.
% Table titles go above tables; all other caption information
%  should be placed in footnotes below the table.
%
%----------------
% EXAMPLE FIGURE
%
 %\begin{figure}
 %\noindent\includegraphics[width=20pc]{samplefigure.eps}
 %\caption{Caption text here}
 %\label{figure_label}
 %\end{figure}
%
% ---------------
% EXAMPLE TABLE
%
%\begin{table}
%\caption{Time of the Transition Between Phase 1 and Phase 2\tablenotemark{a}}
%\centering
%\begin{tabular}{l c}
%\hline
% Run  & Time (min)  \\
%\hline
%  $l1$  & 260   \\
%  $l2$  & 300   \\
%  $l3$  & 340   \\
%  $h1$  & 270   \\
%  $h2$  & 250   \\
%  $h3$  & 380   \\
%  $r1$  & 370   \\
%  $r2$  & 390   \\
%\hline
%\end{tabular}
%\tablenotetext{a}{Footnote text here.}
%\end{table}

% See below for how to make sideways figures or tables.

\end{document}